# Molecule-dynamic-based Aging Clock and Aging Roadmap Forecast with *Sundial*

Wei Wu[1], Zizhen Deng[1], Chi Zhang[1], Can Liao[2], Jinzhuo Wang[1]

1. Department of Big Data and Biomedical AI, College of Future Technology, Peking University, Beijing 100871, China
2. Institute of Bioinformatics, University of Georgia, Athens, GA, United States

Corresponding author:

Jinzhuo Wang, PhD

E-mail: wangjinzhuo@pku.edu.cn

*Sundial, the earliest type of timekeeping device*

**Abstract**

Addressing the unavoidable bias inherent in supervised aging clocks, we introduce *Sundial*, a novel framework that models molecular dynamics through a diffusion field, capturing both the population-level aging process and the individual-level relative aging order. *Sundial* enables unbiased estimation of biological age and the forecast of aging roadmap. Faster-aging individuals from *Sundial* exhibit a higher disease risk compared to those identified from supervised aging clocks. This framework opens new avenues for exploring key topics, including age- and sex-specific aging dynamics and faster yet healthy aging paths.

**Main**

Aging clocks, which use supervised learning to predict biological age (BA) from molecular measurements, have made significant strides in recent years[1–9]. However, a recent controversy has arisen, suggesting that the BA predicted by these supervised aging clocks may be flawed[10]. The core issue lies in the fact that while the goal is to predict BA, chronological age (CA) is often used as the supervisory label. All aging clocks evaluate their accuracy by the degree of alignment between the predicted BA and CA, but if the predicted BA perfectly aligns with CA, the aging clock becomes meaningless. The correlation between predicted BA and CA has been misinterpreted as an indicator of model accuracy, masking a deeper issue: the clock may incorrectly predict misaligned samples as aligned ones. Besides, non-diseased individuals are often assumed to represent cases where BA aligns with CA. This assumption is also problematic, as the absence of disease does not necessarily signify typical aging - neither faster nor slower aging - and BA itself lacks a universally accepted definition[11,12]. The best way to understand a system is by being able to predict its future state. However, aging clocks can only predict the current state of an individual's aging system with bias, because the supervised paradigm merely fits molecular features associated with different BAs, without modeling the dynamics of the entire aging process. We highlight the supervised clock's inherent bias, and argue that aging clocks must evolve beyond the constraints of the CA-supervised framework. We propose that a shift from reductionism to holism may be required to fully understand the aging process.

In this study, we developed *Sundial* (**S**ystematic **Un**biased Molecule-**D**ynamic-**I**nspired **A**ging **L**andscape), a novel framework that modeled molecular dynamics through a diffusion field, capturing both the population-level aging process and the individual-level relative aging order. *Sundial* enabled unbiased estimation of BA through carefully designed distribution mapping (**Methods**). Through the visualization of the transition matrix as a stream, *Sundial* offered an intuitive landscape to observe aging process and uniquely forecasted aging roadmaps. In the following, we outlined the methodology in detail and demonstrated its effectiveness using proteomics data from the UK Biobank[13,14].

The molecular dynamic landscape of aging mirrors the spread of a river, creating a floodplain-like structure **(Fig. 1b).** We constructed a diffusion field[15] in a k-nearest neighbors (KNN) graph to model this structure. Transitions between samples in the graph were interpreted as a diffusion process with the diffusion distance representing the relative difference in BA between samples. Spectral analysis of the transition matrix, via eigenvalue decomposition, enabled the calculation of diffusion distance and mapped samples into a diffusion space. A set of younger samples was selected as the root, and diffusion distances from the root were used to represent the relative aging order, termed pseudo-age. For males (**Extended Data Fig. 1b: middle panel**), the pseudo-age distribution closely mirrored CA, indicating a stable rate of molecular changes. In contrast, females exhibited a bimodal distribution (**Extended Data Fig. 1a: middle panel**), with a pronounced acceleration of molecular changes between ages 45–55, corresponding to menopause. The pseudo-age distributions revealed distinct dynamic patterns in aging across sexes and stages.

To accurately represent an individual's aging degree using the difference between predicted BA and CA, it is essential to ensure that the distributions of these two variables are consistent within the population. To address this, we developed the Age Distribution Back-Mapping rule (**Methods**), which ensured that the distribution of predicted BA aligns with the distribution of CA (**Extended Data Fig. 1a, b: left and right panel**).

We compared the predicted BA from *Sundial* with the predicted BA from the supervised clock (**Methods**) to visually demonstrate the supervised clock's biases and *Sundial*'s advantages. As shown in the 2D visualization of the samples in the middle panel of **Fig. 1c, d,** the color depth represented the CA value, with CA increasing along the horizontal axis. However, within the black box, some younger CA samples appeared on the right, mixing with the older CA group, indicating they are faster-aging samples. Despite this, the supervised clock (**Fig. 1c, d: right panel**) predicted their BA to align with CA, illustrating a clear bias. This occurred because faster-aging individuals are randomly assigned to both the training and test sets, leading to biased BA predictions in the test set. In the left panel of **Fig. 1c, d**, the BA predicted by *Sundial* aligned with the population-level aging trend, allowing abnormal samples to be predicted with a BA greater than their CA, thus identifying them as faster-aging.

In **Extended Data Fig. 2a, c**, we observed a correlation between the predicted BA from *Sundial* and CA, demonstrating that *Sundial* captured aging information. Additionally, the correlation between the predicted BA from *Sundial* and that from the supervised clock further confirmed that *Sundial* captured similar information. It is important to note that the correlation only demonstrated that the clock's understanding of the general aging trend across the population, but it does not ensure precise individual-level predictions. This significant discrepancy between the preliminary BA prediction and CA for certain samples is meaningful, as BA prediction models were typically normalized and age-adjusted during their application (**Extended Data Fig. 2b, d**). What is more crucial for an aging clock is its ability to identify samples with misaligned CA and BA, whereas supervised clocks focus on maximizing the correlation between CA and BA.

To further demonstrate the superior accuracy of *Sundial* in identifying individuals with faster aging, we categorized participants into two groups: Group 1 included those identified as over-aged by *Sundial* but not by the supervised clock, while Group 2 comprised individuals identified as over-aged by the supervised clock but not by *Sundial*. We then compared the risk of eight age-related diseases between these two groups. The results (**Extended Data Fig. 3**) showed that, for all eight diseases, individuals in Group 1 exhibited a higher risk. These findings provided additional evidence of *Sundial*'s enhanced sensitivity in detecting abnormal samples, particularly those with a significant discrepancy between BA and CA.

We visualized the transition matrix as a stream (**Fig. 2a, d**) to forecast the molecular evolution roadmap of individuals (**Fig. 2b, e**). Using k-means clustering, the aging roadmap of females was classified into three groups (0-1, 0-2, 0-3), and similarly, males were categorized into three groups (0-1, 0-2, 0-3). A comparison of disease risks across these roadmap groups showed that females in the (0-2) roadmap and males in the (0-2) roadmap exhibited relatively lower risks for various diseases (**Extended Data Fig. 4**). We then isolated the faster-aging individuals from each roadmap group and performed the same analysis (**Fig. 2b, f**, and **Extended Data Fig. 5c, d**). Faster-aging females in the (0-2) roadmap exhibited significantly lower disease risks compared to those in the other two roadmaps. Similarly, faster-aging males in the (0-2) roadmap had the lowest disease risks, although their risks for certain diseases were comparable to those in the (0-3) roadmap. In contrast, individuals in the (0-1) roadmap had higher risks across most diseases. These results enable *Sundial* to perform subcategorization and disease risk stratification for faster-aging populations.

In summary, our study made three major contributions to the field of aging research. First, we identified and addressed longstanding issues in the construction of aging clocks, including inherent flaws in supervised methods and the misleading nature of current evaluation metrics. Second, we proposed a novel framework that directly models the molecular evolution underlying the aging process, enabling the measurement of the relative order of individual aging and unbiased BA predictions through distribution mapping. Individuals identified as faster aging by *Sundial* demonstrated a higher risk for age-related diseases compared to those predicted by supervised clocks. Furthermore, unlike supervised clocks, which were restricted to evaluating the current state of aging, *Sundial* captured the dynamic molecular progression of aging, offering a direct visualization of aging roadmaps. We showed that distinct aging roadmaps correlated with different disease risks, and that faster-aging individuals along various roadmaps exhibited varied disease risks. *Sundial* escaped the unavoidable biases inherent in supervised clocks, facilitating both BA estimation and aging roadmap prediction. It is concise, effective, and easily applicable to various biological data.

# Methods

## Human cohorts

The UK Biobank is a prospective cohort study with extensive genetic and phenotype data available for 500,000 individuals in the United Kingdom who were recruited between 2006 and 2010. The UKB-PPP is a precompetitive consortium of 13 biopharmaceutical companies funding the generation of blood-based proteomic data from UK Biobank volunteer samples. We restricted our UKB sample to those participants with Olink Explore data available at baseline who were randomly sampled from the main UKB population (n = 53,013).

## Demographic Variables and Chronological Age for Sample

The following demographic variables from the UK Biobank dataset were used to characterize participants: Age at Recruitment (Field ID 21022): Age of the participant at the time of recruitment. Month of Birth (Field ID 52): Birth month of the participant. Year of Birth (Field ID 34): Birth year of the participant. Sex (Field ID 31): Sex of the participant, recorded as male or female. We used the time difference between the date of blood sample collection (Field ID 3166) and the participant's birth month and year to represent the age of each sample. Any samples lacking the necessary information for this calculation were excluded from the analysis.

**Outcomes**
Health outcomes were defined using specific criteria based on initial diagnoses recorded in the UK Biobank: Heart Failure: Identified as the first recorded diagnosis of heart failure (Field ID131354). Kidney Failure: Determined by the initial diagnosis of acute renal failure (Field ID 132030), chronic renal failure (Field ID 132032), or unspecified renal failure (Field ID 132034). Stroke: Defined as the first recorded incidence of stroke, with no specification regarding type (Field ID 131368). Myocardial Infarction: Defined by the first recorded diagnosis of either acute myocardial infarction (field 131298) or subsequent myocardial infarction (Field ID 131300). Chronic Obstructive Pulmonary Disease: Identified as the first recorded occurrence of chronic obstructive pulmonary disease (Field ID 131492). Type 2 Diabetes: Defined as the first diagnosis of non-insulin-dependent diabetes mellitus (Field ID 130708). Liver Cirrhosis/Fibrosis: Defined by the earliest recorded diagnosis of liver fibrosis or cirrhosis (Field ID 131666). Hypertension: Identified by the initial diagnosis of essential (primary) hypertension (Field ID 131286) or secondary hypertension (Field ID 131294). Each condition was categorized based on its first recorded diagnosis to ensure consistency and accuracy in classification across the dataset.

**Age-related Feature Selection**
Feature selection combined Boruta and recursive feature elimination (RFE) to identify proteins predictive of chronological age. Boruta used shadow features as noise benchmarks, iteratively removing features with lower SHAP values until only relevant ones remained. RFE then refined the selection using five-fold cross-validation on the UK Biobank dataset, removing the least important proteins based on SHAP values until five remained. The minimal subset for accurate prediction was determined to be 20 proteins, as reducing this number significantly impaired performance.

**Modeling Molecular Dynamics of Human Aging**
We modeled the aging process as a diffusion field on a graph. Each sample was represented as a node in the graph, and the molecular changes associated with aging were conceptualized as transitions between these nodes. Specifically, the KNN algorithm was used to construct the graph, identifying the k-nearest neighbors for each sample based on the similarity of their age-related molecular profiles. These profiles were carefully selected to ensure the inclusion of informative and non-redundant features, enhancing the reliability of the KNN algorithm in capturing relevant molecular patterns. The constructed graph was analyzed to model the diffusion process, representing molecular evolution during aging.

Transitions between nodes were quantified by their Diffusion Distance, which reflects the relative difference in chronological age between nodes. This distance was computed using a random-walk strategy to capture the transition probabilities between samples. However, this approach can be computationally expensive for large graphs. To address this limitation, we represented the graph as a transition matrix T, where each entry $T_{ij}$ denotes the probability of transitioning from node i to node j. Spectral analysis of the transition matrix provided a more computationally efficient strategy for estimating diffusion distances. By decomposing T into eigenvalues and eigenvectors, the connectivity between nodes could be assessed through differences in their feature vectors, which approximate their diffusion distances. The resulting representation of samples in this eigenvector-derived space is referred to as the Diffusion Space. To quantify molecular aging relative to a reference point, we selected appropriate young samples as the roots of the graph. The pseudo-age for each sample was then computed as the diffusion distance from the root, serving as a proxy for the relative CA difference. This framework offers a robust and interpretable means of modeling and visualizing the molecular dynamics of aging in a graph-based field.

**Age Distribution Back-Mapping**
To convert the relative BA difference represented on a 0–1 scale into a meaningful BA for each individual, we implemented a back-mapping procedure. This approach aligns pseudo-age with human age by leveraging the original CA distribution of the cohort. Rank Assignment of pseudo-age: For each individual, the rank order index_1 of their pseudo-age value was determined within the cohort of n individuals. This step provides a relative ranking based on the pseudo-age scale. Uniform Sampling of Original Age Distribution: The age distribution of the cohort was used to generate n uniformly sampled age values. These sampled values were assigned rank orders index_2 based on their sorted positions in the sampled set. Mapping pseudo-age to Proteomic Age: The pseudo-age rank order index_1 was matched with the rank order of the sampled age distribution index_2. This alignment allowed each individual's pseudo-age (0–1) to be mapped to a corresponding proteomic age. 4. Scaling to Human Age Range: To finalize the mapping, the proteomic age was scaled to span a realistic human age range (40–70 years). This transformation ensures the pseudo-age-derived BA aligns with known CA ranges for the cohort.

**Supervised Aging Clock**
We calculated protein-predicted BA for the filtered cohort using five-fold cross-validation. Within each fold, a Light-GBM model was trained using the tuned hyperparameters and predicted age values were generated for the test set of that fold. We then combined the predicted age values from each of the folds to create a measure of BA for the entire sample.

**Figure 1**

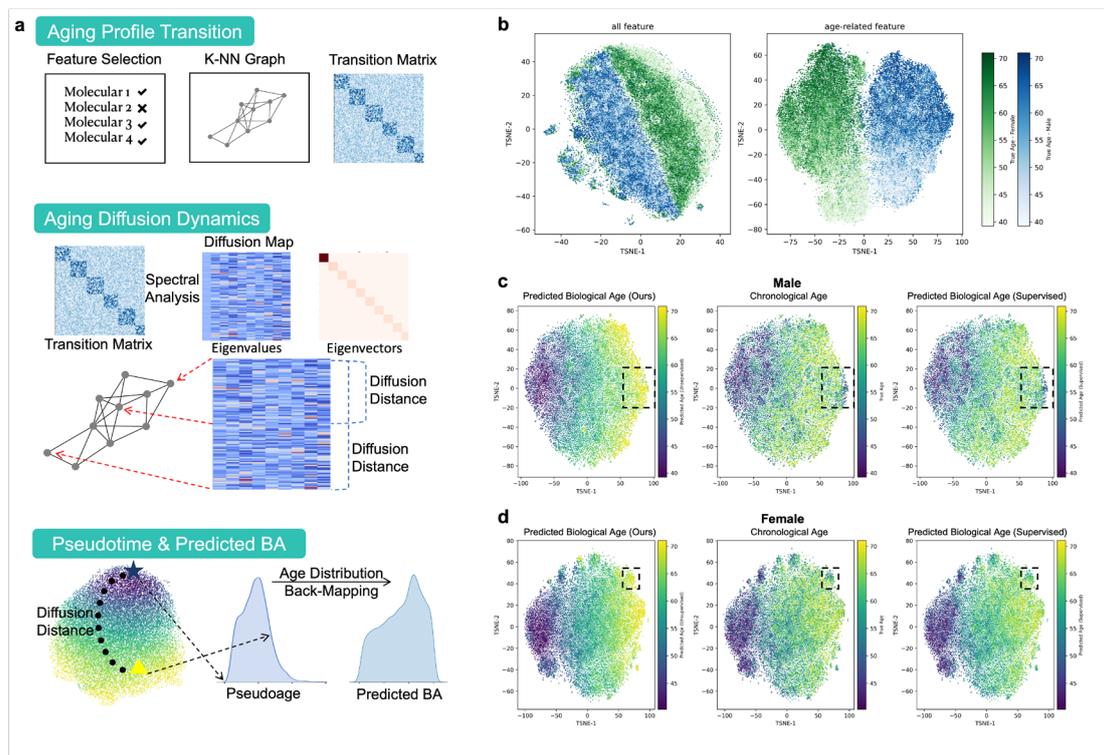

**Figure 1. Workflow of *Sundial* and Comparison of Sundial and Supervised Clock. a.** Overview of the three main steps for *Sundial*: (1) A KNN algorithm was applied to construct a graph based on the similarity of age-related molecular profiles after age-related feature selection. The graph was represented as a transition matrix. (2) Spectral analysis was performed on the transition matrix to embed samples into a diffusion space, where diffusion distance represents the relative BA difference between samples. (3) Distances between all samples and root samples were calculated to generate a pseudo-age, reflecting the relative aging order, which was then mapped to the CA distribution to estimate unbiased BA. **b.** t-SNE visualization for all features and age-related features. Molecular features related to aging were retained, capturing changes such as fluid diffusion. Samples were colored based on sex (male: green; female: blue) and CA, where color depth corresponds to increasing CA. **c, d.** t-SNE visualization for predicted BA (*Sundial*), CA, and supervised predicted BA, shown in the left, middle, and right panels, respectively. Faster-aging individuals are highlighted with bounding boxes. The middle panel indicates individuals with faster aging. The left panel shows that *Sundial* accurately identifies these individuals, while the right panel demonstrates the supervised method's inability to do so effectively.

Figure 2

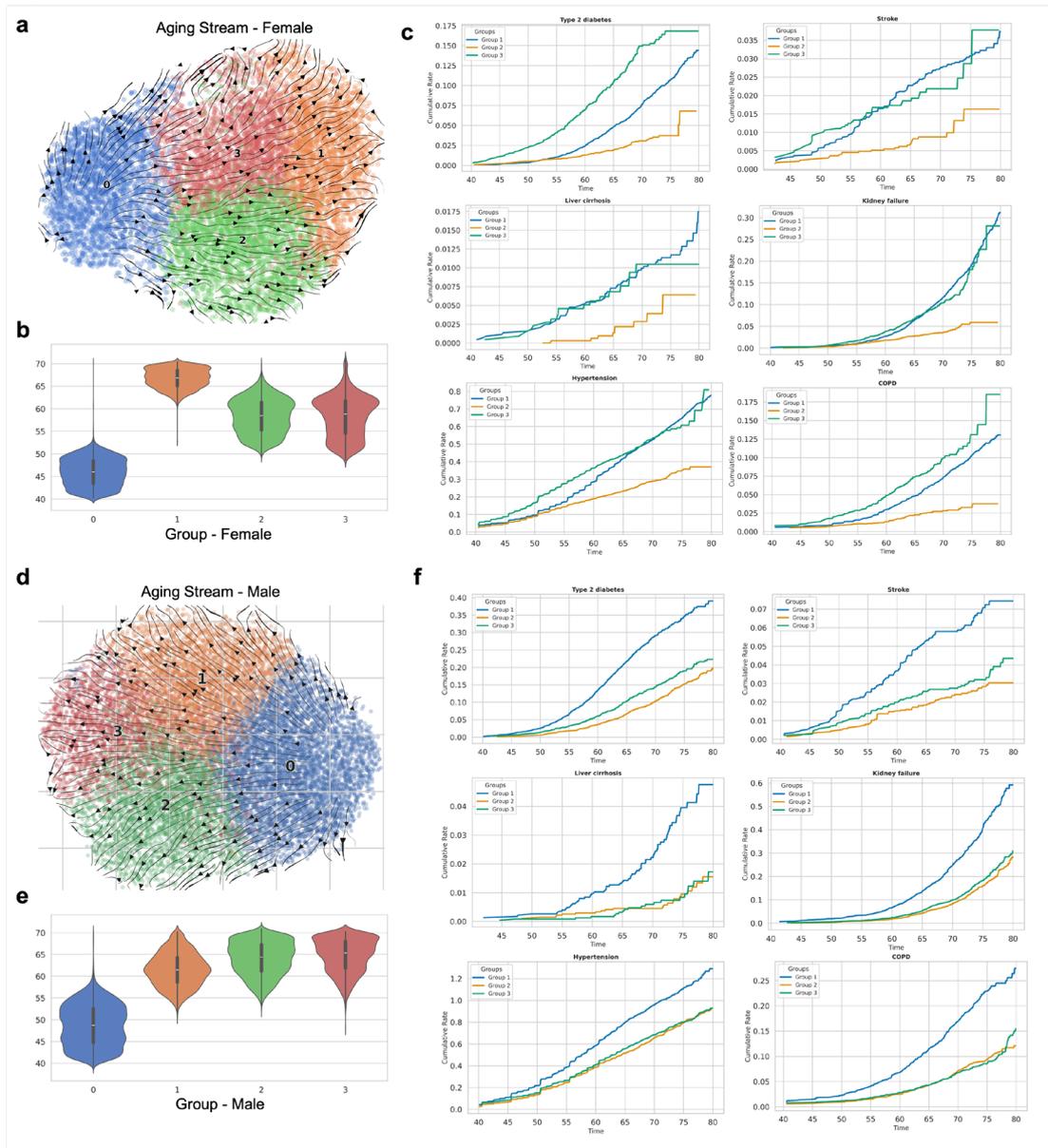

**Figure 2. Aging Roadmaps by Stream and Disease Risk in Different Aging Roadmaps. a,** Visualization of the transition matrix for female by stream, depicting the aging roadmaps. Each line represents a potential transition between molecular profile states, with the direction indicating the aging progression. The color coding corresponds to the distinct aging roadmaps identified through k-means clustering. **b,** Violin plots illustrating the age distribution within each of the three identified female aging roadmaps (0-1, 0-2, 0-3). **c,** Cumulative disease hazard curves for faster-aging (BA-CA>0) female across the three aging roadmaps. (Disease: Type 2 diabetes, stroke, liver cirrhosis, hypertension, lung cancer, and kidney failure). d, e, f, Mirrors panel a, b, c but for male.

**Extended Data Fig. 1**

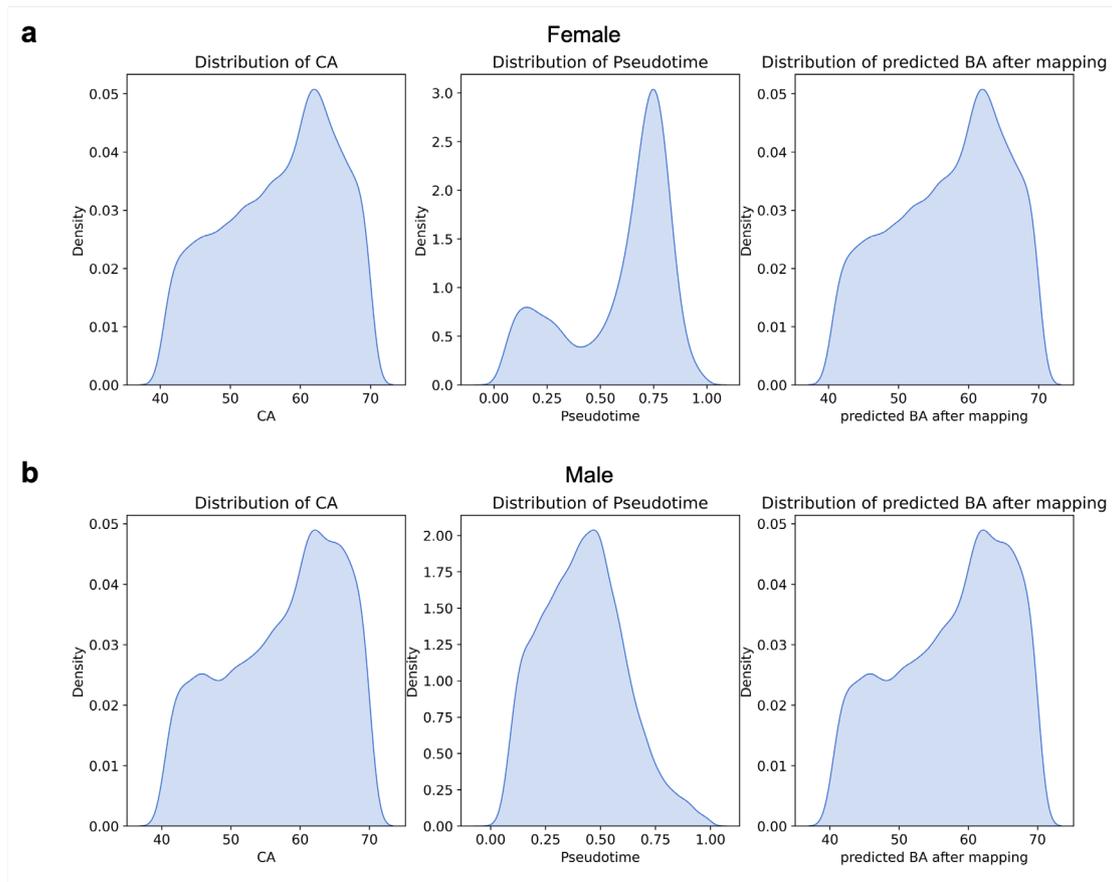

**Extended Data Fig. 1. Age Distribution Back-Mapping. a,** Female distributions for CA (left), pseudo-age (middle), and back-mapped predicted BA (right) are shown. **b,** Male distributions for CA (left), pseudo-age (middle), and back-mapped predicted BA (right) are shown.

**Extended Data Fig. 2**

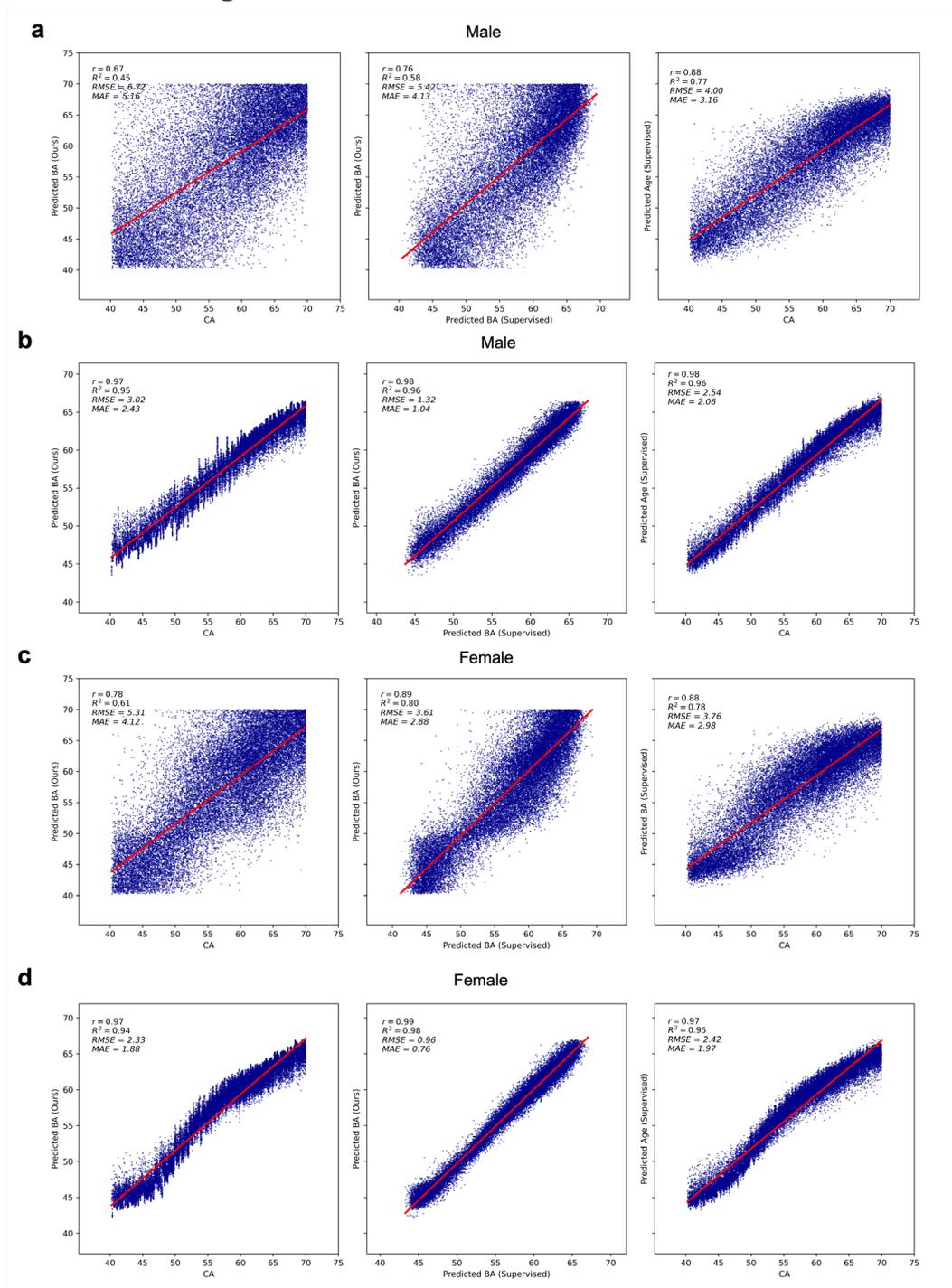

**Extended Data Fig. 2. Performance and Correlation of Aging Clock: a,** Scatter plots focusing on male are presented to evaluate the BA prediction. Left: *Sundial*'s predicted BA against CA. Middle: Sundial's predicted BA against the supervised method's predicted BA. Right: Sundial's predicted BA against CA. **b,** showing the normalized and age-adjusted BA predictions, ensuring the model's applicability. **c,** Scatter plots focusing on female, with the same three comparisons as in panel **a**. **d,** Scatter plots for focusing on female participants, with the same three comparisons as in panel b.

**Extended Data Fig. 3**

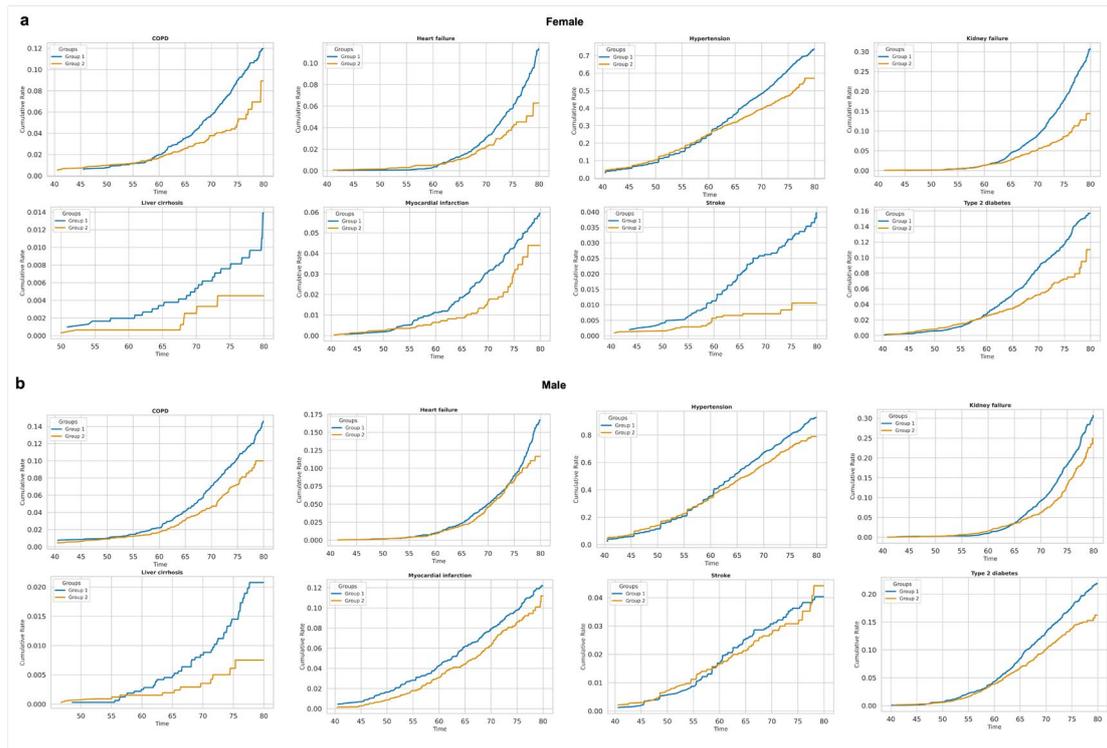

**Extended Data Fig. 3. Comparison of Disease Risk Between Two Faster-aging Groups: a,** Displays the cumulative hazard curves for eight diseases in female, comparing Group 1 (identified as over-aged by *Sundial* but not by the supervised approach, shown in blue) and Group 2 (identified as over-aged by the supervised method but not by *Sundial*, shown in orange). Each sub-panel corresponds to a specific disease, with the y-axis representing the cumulative incidence and the x-axis indicating age. **b,** Mirrors Panel a but for male.

**Extended Data Fig. 4**

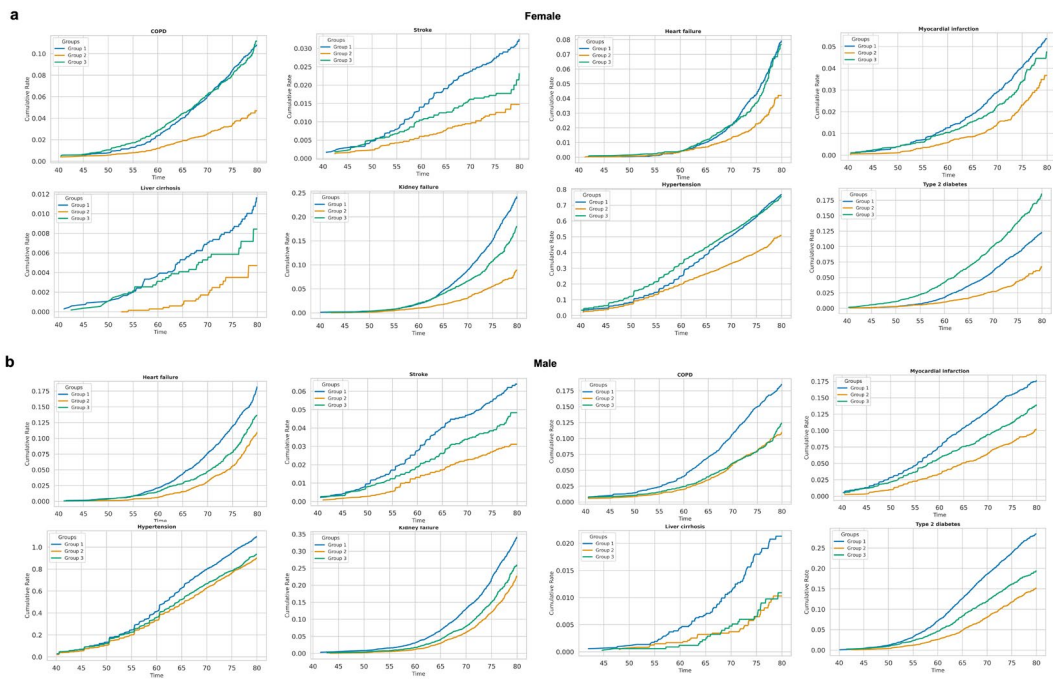

**Extended Data Fig. 4. Comparison of Disease Risk in Different Aging Roadmaps: a,** Displays the cumulative hazard curves for eight diseases in female, across three aging Roadmaps (Groups 1, 2, 3) is show. Each sub-panel corresponds to a specific disease, with the y-axis representing the cumulative incidence and the x-axis indicating age. **b,** Mirrors Panel a but for male.

**Extended Data Fig. 5**

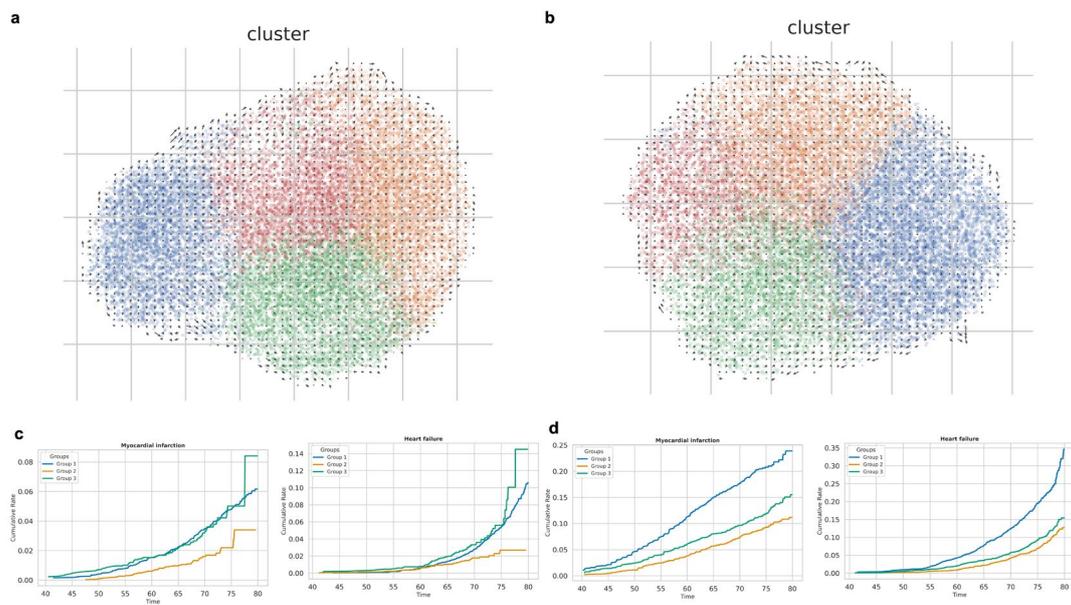

**Extended Data Fig. 5. Aging Roadmaps by Arrows and Disease Risk in Different Aging Roadmaps. a,** Visualization of the transition matrix for female, depicting the molecular aging roadmaps by arrows. Each arrow represents a potential transition between molecular states, with the direction indicating the aging progression. The color coding corresponds to the distinct aging directions identified through k-means clustering. **b,** Mirrors panel a but for male. **c,** Cumulative hazard curves for faster-aging (BA-CA>0) female across the three aging roadmaps (Myocardial Infarction and Heart Failure). **d**, Mirrors panel c but for male.